\documentclass[conference]{IEEEtran}
\IEEEoverridecommandlockouts
\usepackage{cite}
\usepackage{amsmath,amssymb,amsfonts}
\usepackage{algorithmic}
\usepackage{graphicx}
\usepackage{textcomp}
\usepackage{xcolor}
\usepackage{flushend}
\usepackage{booktabs}
\usepackage{subcaption}
\usepackage{makecell}
\usepackage{placeins}

\usepackage{soul}
\usepackage{wrapfig}
\usepackage[caption=false]{subfig}
\usepackage[font=footnotesize,labelfont=bf]{caption}
\usepackage{dblfloatfix}

\usepackage{wasysym}

\usepackage{pifont}
\newcommand{\cmark}{\ding{51}}
\newcommand{\xmark}{\ding{55}}
\newcommand{\pmark}{\LEFTcircle}

\usepackage[a4paper, total={184mm,239mm}]{geometry}
\def\BibTeX{{\rm B\kern-.05em{\sc i\kern-.025em b}\kern-.08em
    T\kern-.1667em\lower.7ex\hbox{E}\kern-.125emX}}

\begin{document}

\title{MeltRTL: Multi-Expert LLMs with Inference-time Intervention for RTL Code Generation}

\author{\IEEEauthorblockN{Nowfel Mashnoor, Mohammad Akyash, Hadi Kamali, Kimia Azar}
\IEEEauthorblockA{\textit{Department of Electrical and Computer Engineering (ECE), University of Central Florida, Orlando, FL 32816, USA} \\
\{nowfel.mashnoor, mohammad.akyash, kamali, azar\}@ucf.edu}
}

\maketitle

\begin{abstract}
The automated generation of hardware register-transfer level (RTL) code with large language models (LLMs) shows promise, yet current solutions struggle to produce syntactically and functionally correct code for complex digital designs. This paper introduces \textit{MeltRTL}, a novel framework that integrates multi-expert attention with inference-time intervention (ITI) to significantly improve LLM-based RTL code generation accuracy without retraining the base model. MeltRTL introduces three key innovations: (1) A multi-expert attention architecture that dynamically routes design specifications to specialized expert networks, enabling targeted reasoning across various hardware categories; (2) An inference-time intervention mechanism that employs non-linear probes to detect and correct hardware-specific inaccuracies during generation; and (3) An efficient intervention framework that selectively operates on expert-specific attention heads with minimal computational overhead. We evaluate MeltRTL on the VerilogEval benchmark, achieving 96\% synthesizability and 60\% functional correctness, compared to the base LLM's 85.3\% and 45.3\%, respectively. These improvements are obtained entirely at inference time, with only 27\% computational overhead and no model fine-tuning, making MeltRTL immediately deployable on existing pre-trained LLMs. Ablation studies further show the complementary benefits of multi-expert architecture and ITI, highlighting their synergistic effects when combined.
\footnote{Code available at: \texttt{https://github.com/mashnoor/melt-rtl}}
\end{abstract}


\section{Introduction}

The integration of large language models (LLMs) into hardware design workflows has sparked growing interest in their ability to translate high-level textual specifications into register-transfer level (RTL) implementations \cite{liu2024rtlcoder, cui2024origen, khan2025sagehls}. By shrinking spec.-to-code time, these models promise faster (and more reliable) prototyping (ultimately a shorter semiconductor time-to-market or TTM) \cite{huang2024towards}. Unlike conventional software generation, however, RTL is a zero-tolerance domain, where designs must precisely capture cycle-accurate behavior while conforming to structural and verification constraints \cite{akyash2025decortl}. In practice, current LLM-generated RTL often compiles but deviates from the specification, e.g., reflecting mismatches, incorrect clocking, missing reset/handshakes, etc. These discrepancies, often categorized as \textit{hallucinations} \cite{ping2025hdlcore}, \cite{yang2025haven}, compromise functional correctness, reliability, and even security \cite{akyash2024evolutionary}.

Reliability in language generation is commonly improved via grounding models with external knowledge \cite{feng2024don}, retrieval-augmented generation \cite{song2024rag}, and tool integration \cite{wu2025combating}. While effective as factual anchors, such methods leave the model's internal computation (e.g., encoding and propagation) unchanged. Recent studies, on the other hand, point to the benefit of directly steering internal representations for truthfulness. Techniques such as activation editing \cite{liao2025qse, qiu2024spectral} and intermediate-level alignment \cite{skean2024does, skean2025layer} demonstrate that errors often arise from systematic misalignments within latent spaces, and by reshaping these dynamics at inference, models can be steered toward truthful and instruction-consistent outputs \cite{zhang2024truthx}.


To the best of our knowledge, no prior work has applied inference-time, activation-level interventions to RTL generation. Extending these advances to hardware is particularly critical: unlike open-domain text, where hallucinations produce factual inaccuracies, hallucinations in RTL manifest as logic that fails simulation or formal checks, degrades power–performance–area (PPA), or undermines security guarantees \cite{ping2025hdlcore, tasnia2025veriopt}. Moreover, the scarcity of high-quality (verified) RTL datasets limits the feasibility of large-scale fine-tuning, making these strategies especially attractive. By steering the internal activations of LLMs during generation, we can directly enforce semantic alignment, ensuring RTL implementations remain reliable, efficient, and instruction-consistent without the prohibitive cost of retraining.

While motivating, leveraging the intermediate layers of LLMs for RTL generation introduces two challenges:

\noindent \textit{\underline{(i) Targeting Representation Components}}: Identifying which elements of the model’s internal representations are most responsible for functional correctness and should be modified. \\
\noindent \textit{\underline{(ii) Robust and Efficient Intervention:}} Designing intervention strategies that are both lightweight and robust, ensuring they effectively enhance RTL functional correctness without introducing instability or excessive computational overhead.

To overcome these challenges, we introduce \textbf{\textit{MeltRTL}}: multi-expert LLMs with inference-time intervention for RTL Code Generation, a framework that couples probe-guided multi-expert attention with inference-time steering, enabling correctness-oriented control without retraining base models. Our contributions are threefold:

\begin{table}[t]
\centering
\scriptsize
\caption{Comparison of Prior LLM-based RTL Generation Models. 
Legend: \cmark~=~yes, \xmark~=~no, \pmark~=~partial/indirect, —~=~not reported.}
\label{tab:rtl_comparison}
\setlength{\tabcolsep}{1.5pt}
\begin{tabular}{lcccccccc}
\toprule
\textbf{Model} & \textbf{Train} & \textbf{Activation} & \textbf{Multi} & \textbf{Agentic} & \textbf{RAG /} & \textbf{Fine} &  \textbf{Open-} \\
& \textbf{free} & \textbf{Steering} & \textbf{Expert} &  \textbf{Tools} & \textbf{Self-Ref.} & \textbf{Tuning} & \textbf{Source}\\
\midrule
RTLCoder \cite{liu2024rtlcoder} & \xmark & \xmark & \xmark & \xmark & \xmark & \cmark FT & \cmark \\
OriGen \cite{cui2024origen} & \xmark & \xmark & \xmark & \pmark  & \cmark & \cmark LoRA FT & \cmark\\
BetterV \cite{pei2024betterv} & \xmark & \xmark & \xmark & \xmark & \pmark  & \cmark FT   & \pmark \\
CodeV \cite{zhao2024codev} & \xmark & \xmark & \xmark & \xmark & \cmark & \cmark FT  & \pmark \\
CraftRTL \cite{liu2024craftrtl} & \xmark & \xmark & \xmark & \xmark & \pmark  & \cmark FT (synt.) & \cmark \\
VerilogCoder \cite{ho2025verilogcoder} & \cmark & \xmark & \xmark & \cmark& \pmark & \xmark  & \xmark \\
MAGE \cite{zhao2024mage} & \cmark & \xmark & \xmark & \cmark & \pmark & \xmark & \cmark  \\
RTL++ \cite{akyash2025rtl++} & \xmark & \xmark & \xmark & \xmark & \pmark  & \cmark FT (graph) & \pmark \\
HDLCoRe \cite{ping2025hdlcore} & \cmark & \xmark & \xmark & \pmark & \cmark & \xmark & — \\
HaVen \cite{yang2025haven} & \xmark & \xmark & \xmark & \xmark & \cmark  & \cmark FT  & \cmark \\
ScaleRTL \cite{deng2025scalertl} & \xmark & \xmark & \xmark & \cmark & — & \cmark (reason) & —\\
VeriSeek \cite{wang2025large} & \xmark & \xmark & \xmark & \xmark & \xmark & \cmark RL & — \\
\midrule
\textbf{MeltRTL} & \cmark & \cmark & \cmark & \pmark & \pmark & \xmark & \cmark \\

\bottomrule
\end{tabular}
\end{table}

\noindent \textit{\underline{\textbf{(i) Probe-Guided Component Identification:}}} We curate a small dataset of 200 samples of instruction–code pairs categorized by hardware specifications (e.g., FSM, arithmetic, etc.). Using this dataset, we train multiple lightweight classifiers on internal activations to identify category-specific attention heads that most strongly predict functional correctness, revealing structure in how LLMs internalize RTL reasoning.

\noindent \textit{\underline{\textbf{(ii) Representation-Level Steering for Correctness:}}} Leveraging the identified heads, MeltRTL applies targeted, non-linear interventions during decoding to enforce specification-faithful behavior. This reduces RTL-specific hallucinations and improves correctness while remaining computationally lightweight (no base model fine-tuning required).

\noindent \textit{\underline{\textbf{(iii) Domain-Specialized Multi-Expert Behavior:}}} We show that steering different subsets of heads enables the model to act as an \textit{expert} in specific hardware design domains. This multi-expert specialization equips MeltRTL to adaptively improve syntactic and functional correctness across diverse categories of RTL designs of varying complexity.

\section{Related Works}

\subsection{LLM for RTL Code Generation}
LLMs show strong capabilities in translating natural language into executable code across software languages, driven by large-scale pretraining on diverse codebases (e.g., Codex \cite{chen2021evaluating}, CodeGen \cite{nijkamp2022codegen}). Recent efforts extend these advances to hardware, covering design optimization \cite{yao2024rtlrewriter}, debugging \cite{yao2024hdldebugger}, and vulnerability detection \cite{mashnoor2025llmift}. Several studies directly target LLM-based RTL generation \cite{wu2024chateda, liu2024craftrtl, zhao2024codev, liu2024rtlcoder, cui2024origen, pei2024betterv}. Early systems \cite{wu2024chateda} rely on prompt engineering and tool feedback (compiler/simulator in the loop) to coax general-purpose LLMs (e.g., GPT-3.5/4) into producing synthesizable code. While helpful, these frameworks are iteration-heavy and often require manual post-processing to ensure semantic fidelity.

To reduce manual intervention, recent research has adapted LLMs specifically for RTL. For instance, RTLCoder \cite{liu2024rtlcoder} augmented scarce datasets by synthesizing instruction–code pairs with GPT-3.5, while OriGen \cite{cui2024origen} introduced iterative refinement through self-reflection and augmentation. BetterV \cite{pei2024betterv} guided generation toward hardware design objectives such as PPA (power, performance, area). CraftRTL \cite{liu2024craftrtl} enriched context with auxiliary artifacts like state diagrams and simulation traces, improving structured reasoning. ScaleRTL \cite{deng2025scalertl} scaled training to a 3.5B-token reasoning dataset that captures RTL semantics and applied reasoning-oriented techniques for RTL code generation. As summarized in Table~\ref{tab:rtl_comparison}, prior efforts have largely focused on three levers: (i) fine-tuning or leveraging large auxiliary datasets, (ii) orchestrating agentic toolchains with compilers and simulators in the loop, or (iii) enriching prompts and contexts with additional artifacts. In contrast, MeltRTL is, to the best of our knowledge, the first approach to directly intervene in the internal representations of LLMs for RTL code generation. This activation-level, train-free strategy is complementary to existing methods and can be readily combined with retrieval-augmented or agentic pipelines.

\begin{figure*}[t]
  \centering
  \includegraphics[width=\linewidth]{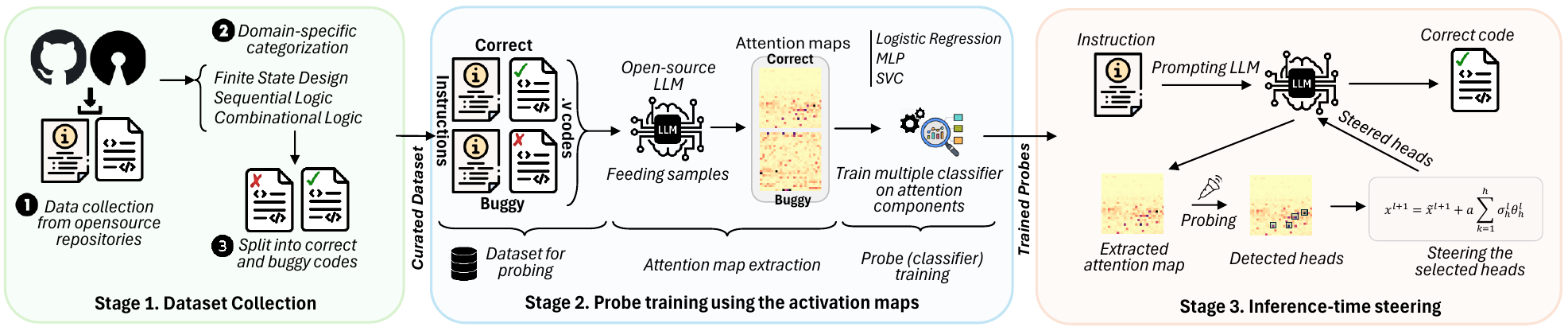}
  \vspace{-15pt}
  \caption{Overall framework of MeltRTL, consisting of three stages: (i) dataset collection, (ii) detection of heads relevant to functional correctness, and (iii) LLM steering via modification of the identified heads.}
  \label{fig:framework}
  \vspace{-15pt}
\end{figure*}

\subsection{Hallucination Mitigation in LLMs}

A growing literature studies internal representations to improve truthfulness and reduce hallucinations \cite{huang2025survey}. While effective in natural language,  direct transfer to structured domains like RTL is nontrivial \cite{li2023iti}, \cite{zhang2024truthx}, \cite{burns2022discovering}.

TruthX \cite{zhang2024truthx} employs a contrastive autoencoder to separate factuality from semantics. Although it shows promise in QA, this method requires finely labeled pairs and incurs lossy reconstruction, which complicates adaptation to RTL tasks. CCS \cite{burns2022discovering} finds truth-related directions through unsupervised contrastive consistency, yet hinges on linguistic negation signals absent in RTL codes.
ITI \cite{li2023iti} ranks attention heads by linear probe accuracy, offering a fast and modular way to attribute truth-related signals. However, its reliance on labeled outputs and the assumption of linear separability limit robustness, and treating heads independently overlooks important joint dependencies. Truth Forest (TrFr) \cite{chen2024truthforest} extends ITI with orthogonal probes and lightweight inference-time steering, improving representation coverage. Still, it remains dependent on labeled data and interpretable latent directions, which restricts its applicability in highly structured and domain-specific settings like hardware generation.

Considering these limitations and the promising results of prior work, MeltRTL is designed as a multi-expert framework for RTL code generation. Unlike existing approaches, MeltRTL directly targets correctness-critical attention heads through probe-guided inference-time steering, reducing hallucinations without requiring extensive labeled data or lossy latent representations. As a result, MeltRTL remains both practical and domain-aware, addressing the strict structural and semantic constraints of hardware description languages.

\section{Proposed Methodology: MeltRTL}

Figure \ref{fig:framework} provides an overview of the MeltRTL framework. As shown, it consists of three main stages: (i) constructing and categorizing an instruction–code dataset (small size), (ii) training probes on intermediate activations to identify correctness-critical heads, and (iii) steering chosen heads at inference time to ensure specification-faithful RTL generation.

\subsection{Dataset Collection for Probe Training}

To enable effective training of the lightweight probes used in MeltRTL, we curate a small dataset of 200 samples comprising paired design specifications (instructions) and corresponding RTL code implementations. The dataset was collected from a wide range of open-source RTL (.v/.sv) repositories, including GitHub, OpenCores, and other community-maintained RTL repositories (examples spanning arithmetic units, controllers, encoders/decoders, and memory components). A key difference between this dataset preparation and the dataset used for fine-tuning base models is its intentionally relatively small size\footnote{Probes are lightweight classifiers/regressors trained on model activations; they need clean, labeled signals more than volume.}. From these repositories, we curated a total of \textbf{200 representative samples}, each consisting of an instruction–code pair. For every RTL module, a corresponding textual specification or derived instruction was either directly obtained, formulated from documentation and code comments, or generated by an LLM (validated manually for correctness and accuracy). To ensure high-quality supervision, we apply a two-stage filtering pipeline: (i) We applied syntactic validation to remove incomplete or ill-formed modules; and (ii) We employed simulation-based checks and assertion-driven formal verification to assess functional behavior. Based on these evaluations, the dataset is partitioned into two categories:

\noindent \textbf{\underline{\textit{(i) Functionally Correct Code:}}} RTL codes that compile and satisfy their design intent without observable deviations.

\noindent \textbf{\underline{\textit{(ii) Functionally Incorrect (Buggy) Code:}}} Modules that are syntactically correct but functionally misaligned, exhibiting errors such as incorrect state transitions, misaligned outputs, or signals with incorrectly initialized values. 

Beyond correctness, we further categorize designs into three fundamental logic classes (w.r.t. their application):

\noindent \textbf{\underline{\textit{(1) Combinational Modules:}}} Stateless modules (w/o register and memory) where outputs depend solely on current inputs.

\noindent \textbf{\underline{\textit{(2) Sequential (Datapath) Modules:}}} Modules driven by registers or memory elements, especially for building datapath (e.g., datapath in pipeline or systolic arrays). 

\noindent \textbf{\underline{\textit{(3) Finite-State Design (Controller):}}} Modules dominated by explicit state machine transitions or controlling signals.

This additional categorization was motivated by our goal of understanding \textbf{which attention heads are critical for which category of design}. By associating probe predictions with functional domains, MeltRTL learns to allocate specialized expertise across different logic types. This categorization directly informed the multi-expert design of our framework, enabling expert-specific intervention strategies that improve generation fidelity across diverse RTL patterns.

\subsection{Classifier (Probe) Training}
\label{sec:probe-training}

To train probes that can detect correctness-related signals, we followed a structured pipeline. First, the curated instruction–code pairs were fed into the pre-trained LLM, and we collected the corresponding activation outputs from individual attention heads. From these activations, we isolated the outputs of individual attention heads, where each head provides a representation vector $h_{i} \in \mathbb{R}^d$. These vectors encode information about intermediate reasoning steps during code generation.  

Next, we labeled the head representations according to the functional correctness of the generated code (linking activations to functional outcomes). Head activations originating from modules classified as correct were assigned label $y=1$, while those from buggy modules were labeled $y=0$. This transformation provided a training dataset in the form $\{(h_{i}, y)\}$, suitable for supervised probe training (probes correlate attention-head activity with RTL correctness).  

\subsubsection{Probe Architectures}\label{subsubsec:probes}

To capture discriminative signals across representations, we trained multiple classifiers: 

\noindent \textbf{\underline{\textit{(i) Logistic Regression (LR):}}} A linear baseline that tests whether correctness can be separated with a hyperplane.

\noindent \textbf{\underline{\textit{(ii) Multi-Layer Perceptron (MLP):}}} A shallow non-linear network that captures higher-order interactions missed by LR.

\noindent \textbf{\underline{\textit{(iii) Support Vector Classifier (SVC):}}} A (RBF) kernel-based margin classifier, useful for separating correctness signals in curved or irregular regions of the space.

Formally, the objective of probe is to learn a specific mapping ($f: h_{i} \mapsto \{0,1\}$), where $0$ denotes buggy code and $1$ denotes functionally correct code. This formalizes the probe’s task as binary classification over head representations. Now, as shown in Equations \ref{eq:lr}, \ref{eq:mlp}, \ref{eq:svc}, the model estimates the probability of correctness based on LR (followed by a sigmoid, which is effective when correctness signals align linearly in feature space), MLP (e.g., ReLU allowing the probe to model interactions across features, capturing more subtle correctness cues), and SVC (carving out complex decision boundaries, making it suitable when correctness signals are not linearly separable.) probes, respectively. 

\begin{equation}
\label{eq:lr}
P(y=1 \mid h_{i}) = \sigma(W^\top h_{i} + b).
\end{equation}
\begin{equation}
\label{eq:mlp}
P(y=1 \mid h_{i}) = \sigma\left(W_2 \, \phi(W_1 h_{i} + b_1) + b_2\right).
\end{equation}
\begin{equation}
\label{eq:svc}
f(h_{i}) = \text{sign}\left(\sum_{j=1}^{m} \alpha_j y_j K(h_{i}, h_{j}) + b\right).
\end{equation}

\subsubsection{Ensemble and Majority Voting}

Each classifier was trained independently, and their predictions were combined into an ensemble using majority voting (Equation \ref{eq:vote}):

\begin{equation}
\label{eq:vote}
\hat{y} = \text{mode}\{ f_{\text{LR}}(h_{i}), f_{\text{MLP}}(h_{i}), f_{\text{SVC}}(h_{i}) \}.
\end{equation}

This ensemble improves robustness by pooling the complementary strengths of linear and non-linear models, reducing sensitivity to the weaknesses of any single probe. Beyond classification, the probes highlight \textbf{which attention heads consistently correlate with correctness}. These correctness-critical heads form the foundation of MeltRTL’s inference-time intervention, where steering is applied selectively to maximize alignment with design intent while keeping overhead low.

\subsection{Inference-Time Steering of Selected Heads}

With a ranked set of correctness-critical heads, inference-time steering can be accomplished in three stages: (i) background formulation of multi-head attention, (ii) probe-guided selection of correctness-critical heads, and (iii) targeted residual corrections applied only to those heads during generation.

\noindent \textbf{\underline{\textit{(i) Attention background:}}} In a transformer layer $l$, each head $h \in \{1,\dots,H\}$ computes $z_h^{\,l}$ as follows:
\begin{equation}
z_h^{\,l} = \mathrm{Att}_h^{\,l}\!\big(P_h^{\,l} x^{\,l}\big),
\end{equation}
where $x^{\,l}\in\mathbb{R}^{d}$ is the residual state, $P_h^{\,l}$ the input projection, and $\mathrm{Att}_h^{\,l}$ the attention function. The heads are projected and aggregated to update the residual stream:
\begin{equation}
\tilde{x}^{\,l+1} = x^{\,l} + \sum_{h=1}^H Q_h^{\,l} z_h^{\,l},
\label{eq:mh-agg}
\end{equation}
with $Q_h^{\,l}$ denoting the output projection. It defines how information from heads is fused into the model’s hidden state.  

\noindent \textbf{\underline{\textit{(ii) Probe-guided head selection:}}} At inference, the ensemble decision from Section~\ref{sec:probe-training} determines whether a head is correctness-critical. Specifically, for head $h$ in layer $l$, the binary gating variable $\sigma_h^{\,l} \in \{0,1\}$ is set by the probe ensemble, and the steering direction $\theta_h^{\,l}\in\mathbb{R}^d$ is a unit-normalized correction vector (nudging activations toward the “functionally correct” region of representation space).  

\noindent \textbf{\underline{\textit{(iii) Inference-time steering:}}} During generation, the residual stream is selectively corrected as:
\begin{equation}
x^{\,l+1} = \tilde{x}^{\,l+1} + \alpha \sum_{h=1}^H \sigma_h^{\,l}\,\theta_h^{\,l},
\label{eq:steer-final}
\end{equation}
where $\alpha>0$ controls intervention strength. This ensures that only correctness-critical heads contribute adjustments, while all other heads remain untouched.  

By design, this mechanism ensures (i) \textbf{targeted interventions}, where only heads identified as correctness-critical are adjusted, and (ii) \textbf{compatibility}, since the additive corrections are lightweight and preserve the model’s overall distribution while nudging it away from error-prone regions.

\section{Results and Evaluation}

\subsection{Experimental Setup}

\begin{table}[t]
\centering
\small
\setlength\tabcolsep{6pt}
\caption{Hyperparameter Configuration. \textit{Full Grid Search over all Parameters with 297 Independent Experiments.}}
\label{tab:hyperparams}
\begin{tabular}{@{} l l @{}}
\toprule
\textbf{Hyperparameter} & \textbf{Values Explored} \\
\cmidrule(r){1-1} \cmidrule(r){2-2}
Probe Type              & Linear (LR), Multi-Layer Perceptron (MLP),\\                            & Support Vector (SVC) with RBF Kernel \\
\cmidrule(r){1-1} \cmidrule(r){2-2}
Top-K Heads             & 5, 10, 15, 20, 25, 30, 35, 40, 45, 48 \\
\cmidrule(r){1-1} \cmidrule(r){2-2}
Intervention Strength ($\alpha$) & 0.1, 0.5, 1.0, 1.5, 2.0, 2.5, \\
                                 & 3.0, 3.5, 4.0, 4.5, 5.0 \\
\cmidrule(r){1-1} \cmidrule(r){2-2}
Probe Max Iterations    & 1000 \\
\cmidrule(r){1-1} \cmidrule(r){2-2}
Probe Regularization    & 0.01 \\

\bottomrule
\end{tabular}
\end{table}

We ran all experiments on a GPU server with clustered NVIDIA H100 GPUs for large-scale training and evaluation. The stack included PyTorch with HuggingFace Transformers for model operations and probing, iVerilog for RTL simulation, and Yosys for RTL synthesizability analysis. The setup involved ~300 runs across multiple parameter dimensions. Probe architectures included logistic regression, MLPs, and SVCs with RBF kernels, implemented in R. Top-K head selection ranged from 5–48, and intervention strength from 0.1–5.0. All probes used 1000 iterations with 0.01 regularization. Hyperparameter details are summarized in Table \ref{tab:hyperparams}.

\subsection{Overall Performance Comparison}

Our results show MeltRTL substantially improves both synthesizability and functional correctness on the VerilogEval Dataset \cite{liu2023verilogeval}. Table \ref{tab:method_comparison} compares the baseline LLM (QwenCoder2.5-14B), MeltRTL (single-expert), and MeltRTL (multi-expert). The baseline reached 85.33\% synthesizability, MeltRTL (single) improved to 93.33\%, and MeltRTL (multi) achieved 96\%. Functional correctness showed larger gains. The baseline achieved 45.33\%, MeltRTL (single) reached 52\%, and MeltRTL (multi) 60\%. As shown in Table \ref{tab:method_comparison}, statistical analysis shows that the multi-expert variant was significant (p = 0.015, d = 0.28), confirming robust improvements, while the single-expert showed less reliable trends.

\begin{figure*}
\centering
\begin{subfigure}{0.32\textwidth}
\includegraphics[width=\linewidth]{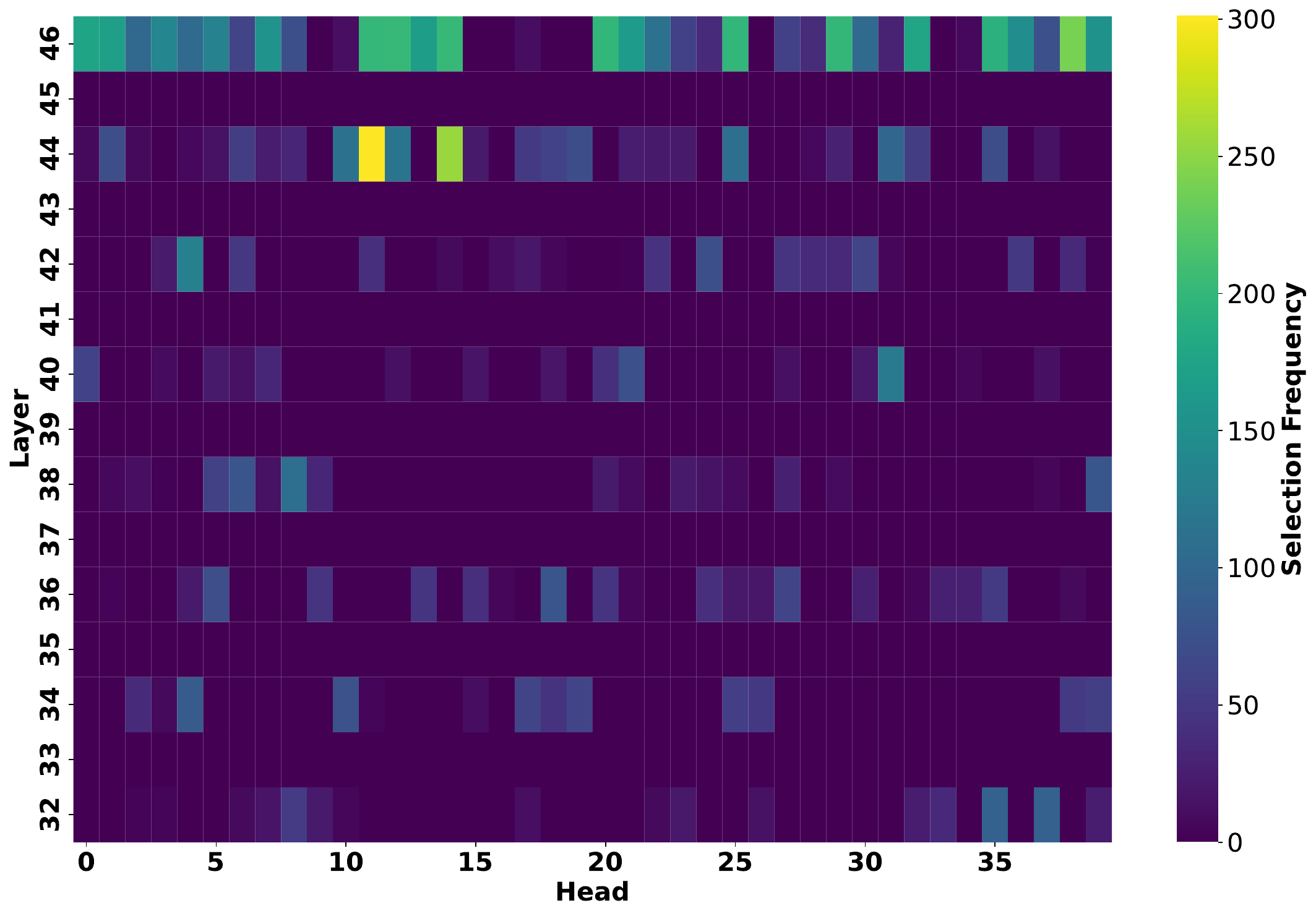}

\caption{Combinational Logic}
\end{subfigure}
\begin{subfigure}{0.32\textwidth}
\includegraphics[width=\linewidth]{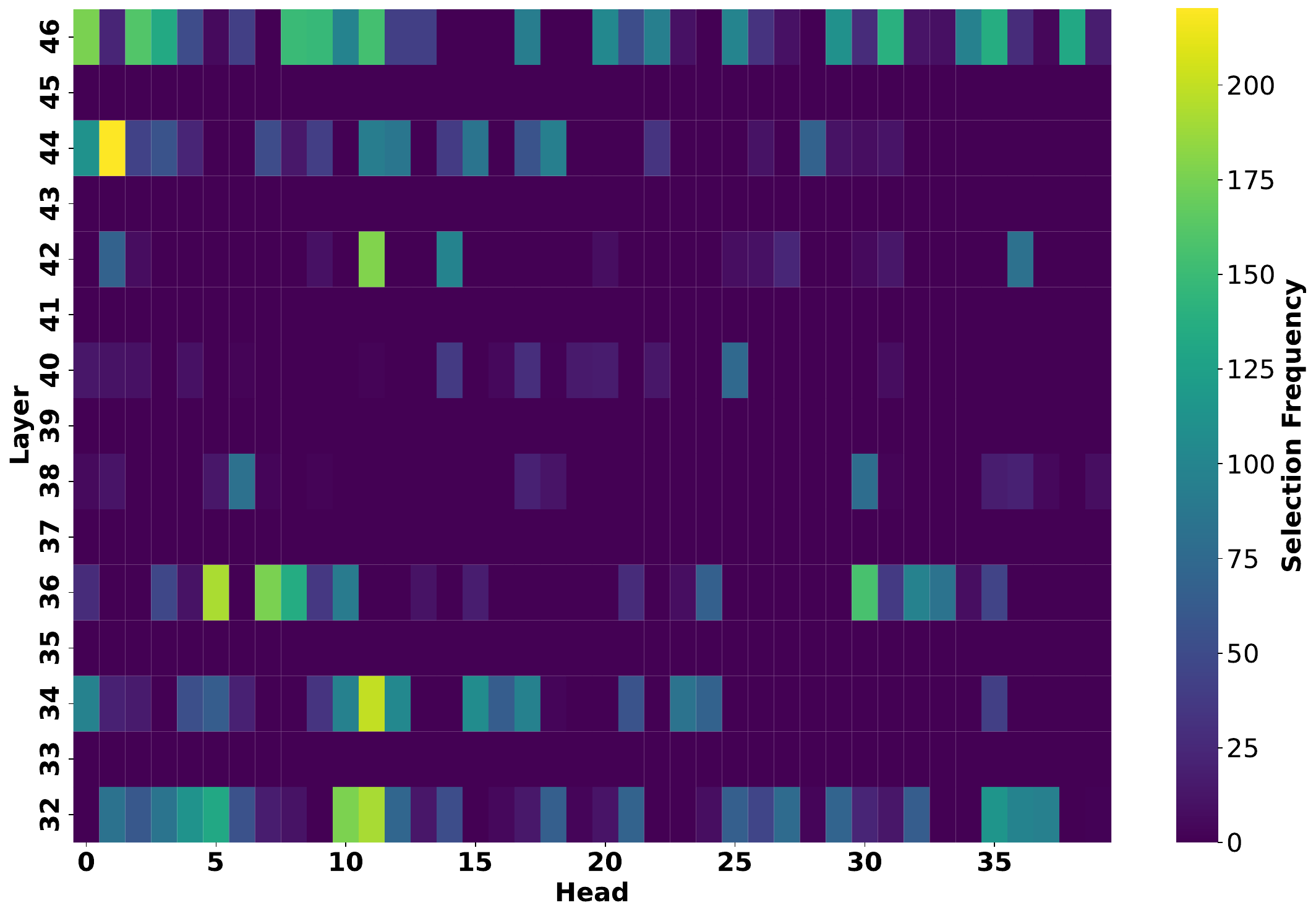}
\caption{FSM (Controller)}
\end{subfigure}
\begin{subfigure}{0.32\textwidth}
\includegraphics[width=\linewidth]{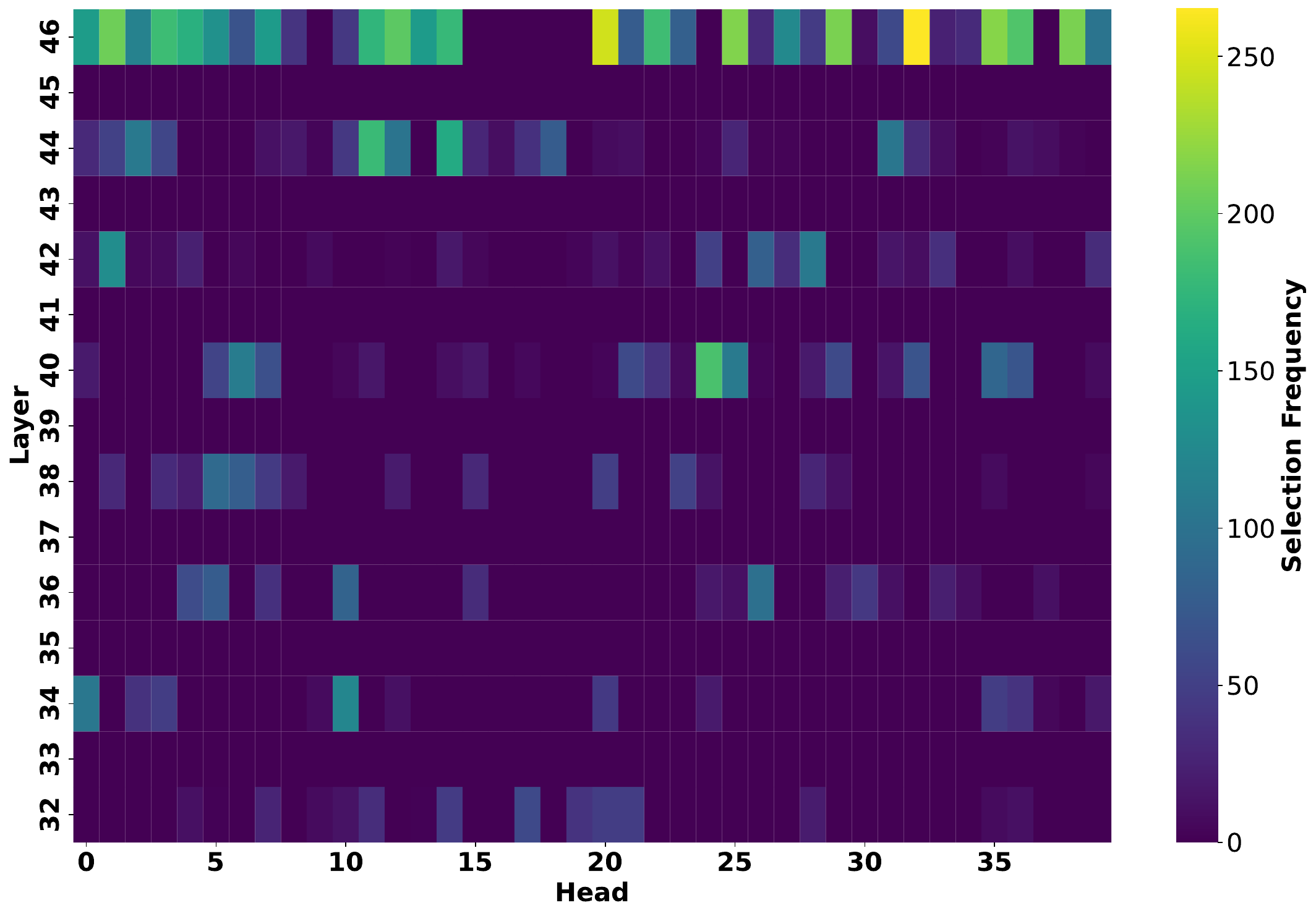}
\caption{Sequential (Datapath) Logic}
\end{subfigure}
\caption{Heatmaps of Correctness-Critical Attention Heads Selection Across Design Classes. Brighter regions indicate heads more strongly correlated with functional correctness, revealing that distinct subsets of heads specialize by design category}
\label{fig:feature_discovery}
  \vspace{-10pt}
\end{figure*}

\begin{table}[b]
\centering
\setlength\tabcolsep{8pt}
\caption{Comparison of Models (Synthesizability and Functional).}
\label{tab:method_comparison}
\begin{tabular}{@{} l c c c c c @{}}
\toprule
\textbf{Method} & \textbf{Synth.} & \textbf{Func.} & \makecell{Impr.\\vs. Base} & \makecell{p-\\val} & \textbf{d} \\
\cmidrule(r){1-1} \cmidrule(r){2-2} \cmidrule(r){3-3} \cmidrule(r){4-4} \cmidrule(r){5-5} \cmidrule(r){6-6}
Base & 85.33\% & 45.33\% & - & - & - \\
MeltRTL (Single) & 93.33\% & 52.00\% & +6.67\% & 0.299 & 0.12 \\
MeltRTL (Multi) & \textbf{\underline{96.00\%}} & \textbf{\underline{60.00\%}} & \textbf{\underline{+14.67\%}} & \textbf{\underline{0.015}} & \textbf{\underline{0.28}} \\
\bottomrule
\end{tabular}
\end{table}

\begin{table}[b]
\centering
\small
\setlength\tabcolsep{3pt}
\caption{Functional Correctness \% by Expert. Columns correspond to evaluated models, rows to problem types. Base: QwenCoder2.5-14B w/o steering. General: MeltRTL with non-specific (global) steering. Comb./Seq./FSM: MeltRTL using single-expert heads for comb., seq., or FSM designs. Multi-Expert: MeltRTL leveraging heads from all three categories jointly.} 
\label{tab:expert_performance}
\begin{tabular}{@{} l c c c c c c @{}}
\toprule
\textbf{Category} & \textbf{Base} & \textbf{General} & \textbf{Comb.} & \textbf{Seq.} & \textbf{FSM} & \textbf{Multi-Expert} \\
\cmidrule(r){1-1} \cmidrule(r){2-2} \cmidrule(r){3-3} \cmidrule(r){4-4} \cmidrule(r){5-5} \cmidrule(r){6-6} \cmidrule(r){7-7}
Comb.  & 68.0\% & 70.7\% & \textbf{\underline{72.7\%}} & 71.0\% & 64.0\% & \textbf{\underline{79.8\%}}\\
Seq.   & 31.0\% & \textbf{\underline{33.3\%}} & 28.6\% & \textbf{\underline{33.3\%}} & 26.2\% & \textbf{\underline{43.6\%}} \\
FSM    & 12.1\% & 6.1\%  & 6.1\% & 15.2\% & \textbf{\underline{15.3\%}} & \textbf{\underline{21.7\%}} \\
\bottomrule
\end{tabular}
\end{table}

\subsection{Expert-Specific Analysis}

To examine MeltRTL’s adaptability, we analyzed performance across three expert classes: combinational logic, sequential (datapath) logic, and FSMs (controllers). Table \ref{tab:expert_performance} compares functional correctness when various categories are used for attention heads and steering. As shown, combinational logic provides the highest baseline at 68\%, improving to 72.7\% with MeltRTL using the combinational expert. Sequential (datapath) logic and FSMs remained difficult, with the base model lowest overall. FSMs were most challenging, with only 12.1\% baseline correctness, but MeltRTL’s multi-expert method improved it by approximately 10\%.

Table \ref{tab:expert_performance} details these metrics, confirming that domain-matched experts yield consistent gains, while the multi-expert model provides the best results, showcasing the importance of category-based attention head and steering. Improvements scale with design complexity, showing specialized interventions are most beneficial for difficult categories.

\subsection{Probe Performance and Head Selection Analysis}

\begin{table}[b]
\centering
\small
\setlength\tabcolsep{7pt}
\caption{Top and Bottom 3 by Functionality and Synthesizability}
\label{tab:leaderboard_combined}
\begin{tabular}{@{} l l c c c c @{}}
\toprule
\textbf{Rank} & \textbf{Cls.} & \textbf{K} & \textbf{Str.} & \textbf{Metric} & \textbf{Score} \\
\cmidrule(r){1-1} \cmidrule(r){2-2} \cmidrule(r){3-3} \cmidrule(r){4-4} \cmidrule(r){5-5} \cmidrule(r){6-6}
Top 1 & SVC with RBF & 15 & 3.0 & Func.  & 60.00 \\
Top 2 & SVC with RBF & 35 & 4.5 & Func.  & 59.13 \\
Top 3 & LR (Linear) & 48 & 0.1 & Func.  & 59.33 \\
\midrule
Top 1 & Multi-Layer (MLP)     & 10 & 2.0 & Synth. & 99.33 \\
Top 2 & LR (Linear)  & 45 & 3.5 & Synth. & 99.33 \\
Top 3 & LR (Linear)  &  5 & 1.5 & Synth. & 98.67 \\
\midrule
Bottom 3 & LR (Linear)  & 15 & 4.0 & Func.  & 37.67 \\
Bottom 2 & LR (Linear)  & 30 & 1.0 & Func.  & 35.22 \\
Bottom 1 & Multi-Layer (MLP)     & 30 & 4.0 & Func.  & 33.67 \\
\midrule
Bottom 3 & Multi-Layer (MLP)     & 15 & 4.5 & Synth. & 81.33 \\
Bottom 2 & Multi-Layer (MLP)     & 48 & 2.5 & Synth. & 79.67 \\
Bottom 1 & SVC with RBF & 45 & 3.0 & Synth. & 74.67 \\
\bottomrule
\end{tabular}
\end{table}

\begin{table}[t]
\centering
\small
\setlength\tabcolsep{8pt}
\caption{Percentage Distribution of Selected Expert Heads in Mid Layers (32--40) vs. Late Layers (42--46) (Expert heads consistently localize in the late (final) layers, especially L44 and L46).}
\label{tab:layer_distribution}
\begin{tabular}{@{} l l c c @{}}
\toprule
\textbf{Model} & \textbf{Expert} & \textbf{Mid (32--40)} & \textbf{Late (42--46)} \\
\cmidrule(r){1-1} \cmidrule(r){2-2} \cmidrule(r){3-3} \cmidrule(r){4-4}
SVC with RBF & General   & 11.6 & 88.4 \\
         & Comb.     & 10.1 & 89.9 \\
         & Seq.      & 16.9 & 83.1 \\
         & FSM       & 26.3 & 73.7 \\
\cmidrule(r){1-1} \cmidrule(r){2-2} \cmidrule(r){3-3} \cmidrule(r){4-4}
LR (Linear)   & General   &  6.5 & 93.5 \\
         & Comb.     &  5.3 & 94.7 \\
         & Seq.      &  7.9 & 92.1 \\
         & FSM       & 22.8 & 77.2 \\
\cmidrule(r){1-1} \cmidrule(r){2-2} \cmidrule(r){3-3} \cmidrule(r){4-4}
Multi-Layer (MLP)      & General   & 53.6 & 46.4 \\
         & Comb.     & 54.7 & 45.3 \\
         & Seq.      & 60.8 & 39.2 \\
         & FSM       & 64.9 & 35.1 \\
\bottomrule
\end{tabular}
\end{table}

To assess the effectiveness of probe-guided head selection (Section \ref{subsubsec:probes}), we analyzed probe performance and classifier discriminability across RTL design experts. As shown in Table \ref{tab:leaderboard_combined}, we tested three probe architectures. For functional correctness, SVC with RBF consistently led, peaking at K=15 and $\alpha$=3.0. LR (linear) probes showed high variance (best 59.33\%, worst 33.67\%), while Multi-Layer Perceptron (MLP) achieved intermediate, more stable results. Figure \ref{fig:feature_discovery} reflects domain-specific head activation: combinational logic concentrated in specific clusters, while sequential logic and FSMs showed more distributed patterns, reflecting higher complexity and need for diverse attention. Table \ref{tab:layer_distribution} shows most expert heads (73.7–94.7\%) localize in layers 42–46 across probes and domains. This trend is strongest for LR and SVC probes, while MLPs distribute more evenly, suggesting non-linear probes can also capture earlier-layer signals relevant to correctness.

\subsection{Activation Space and Discriminative Analysis}

\begin{figure*}
\centering
\begin{subfigure}{0.325\textwidth}
\includegraphics[width=\linewidth]{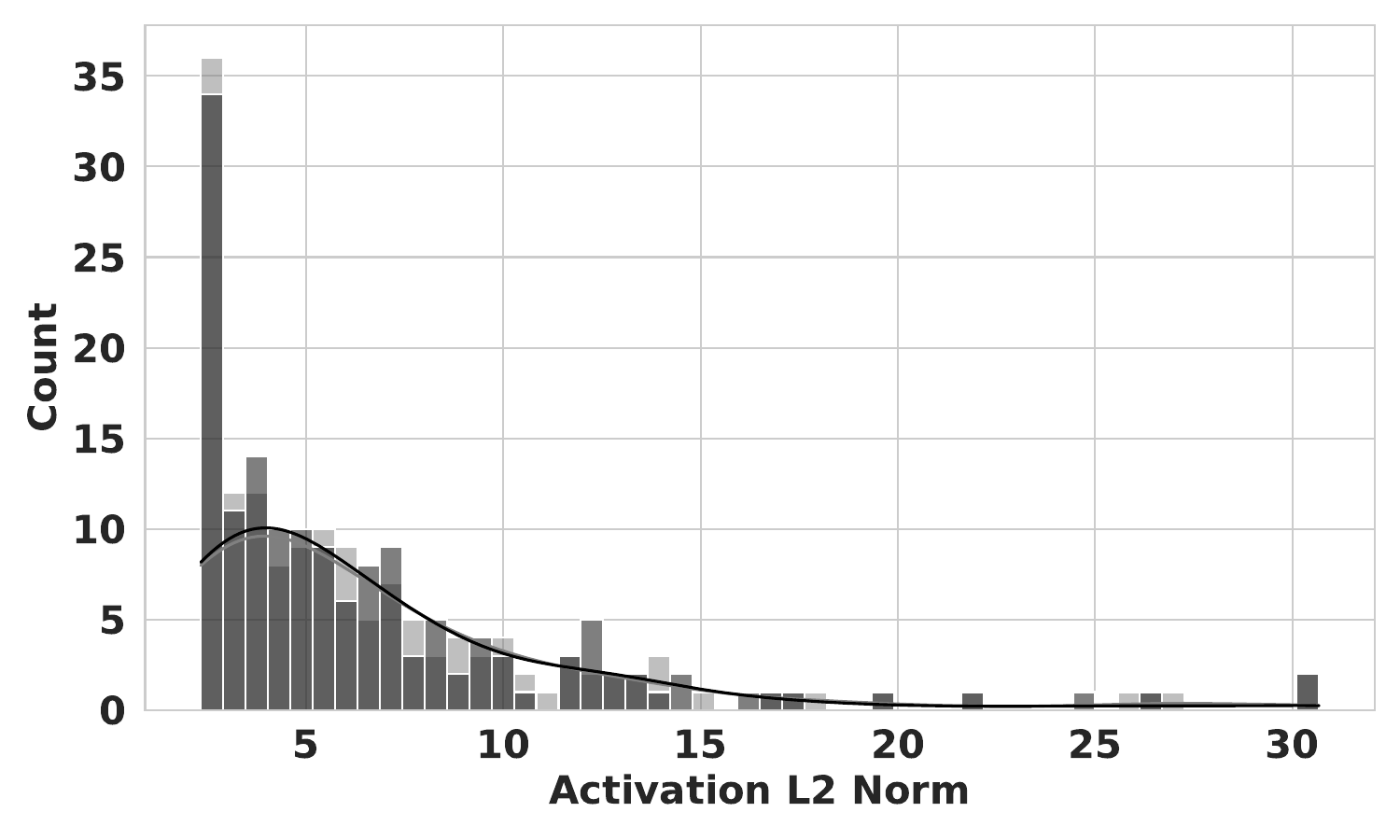}
\caption{SVC with RBF}
\end{subfigure}
\begin{subfigure}{0.325\textwidth}
\includegraphics[width=\linewidth]{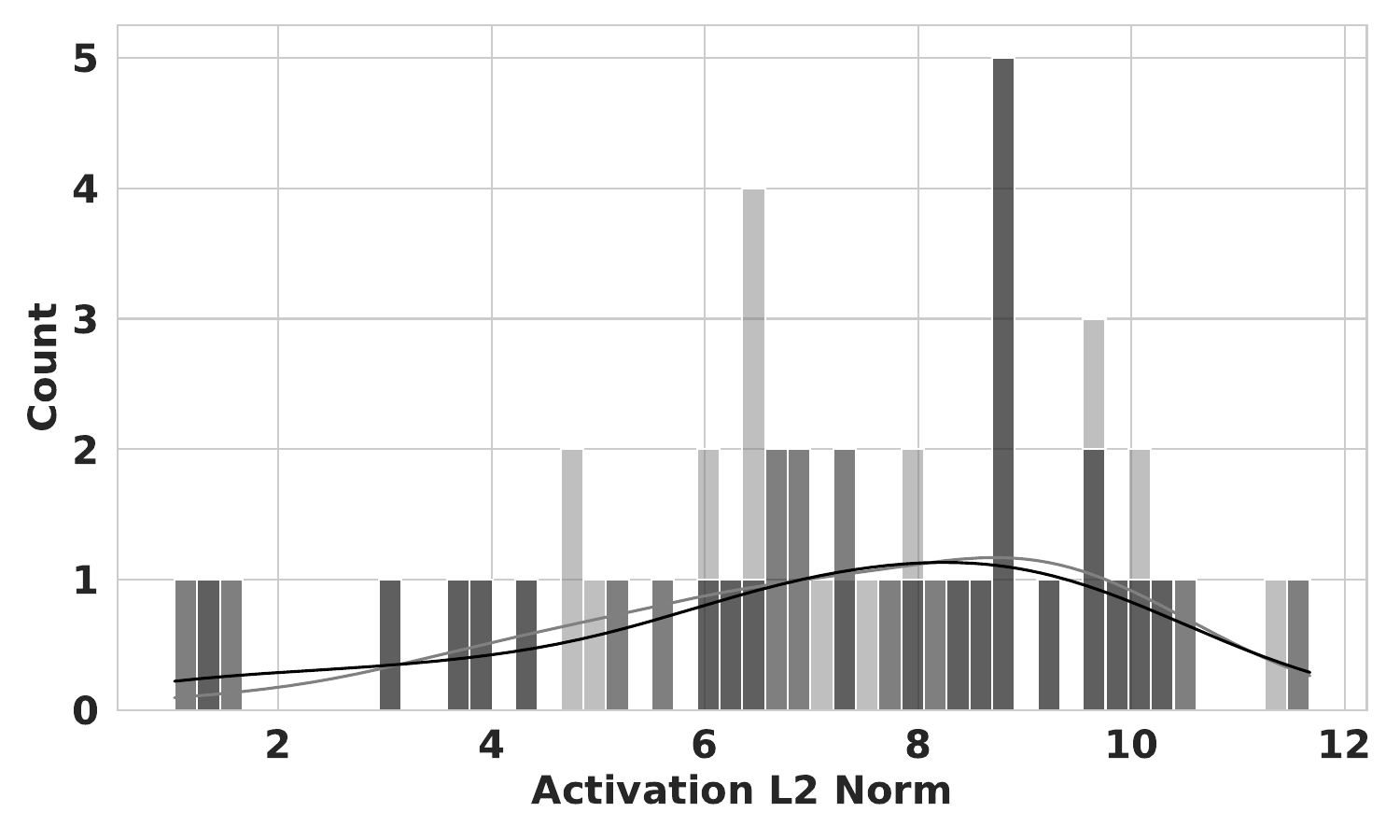}
\caption{LR (Linear)}
\end{subfigure}
\begin{subfigure}{0.325\textwidth}
\includegraphics[width=\linewidth]{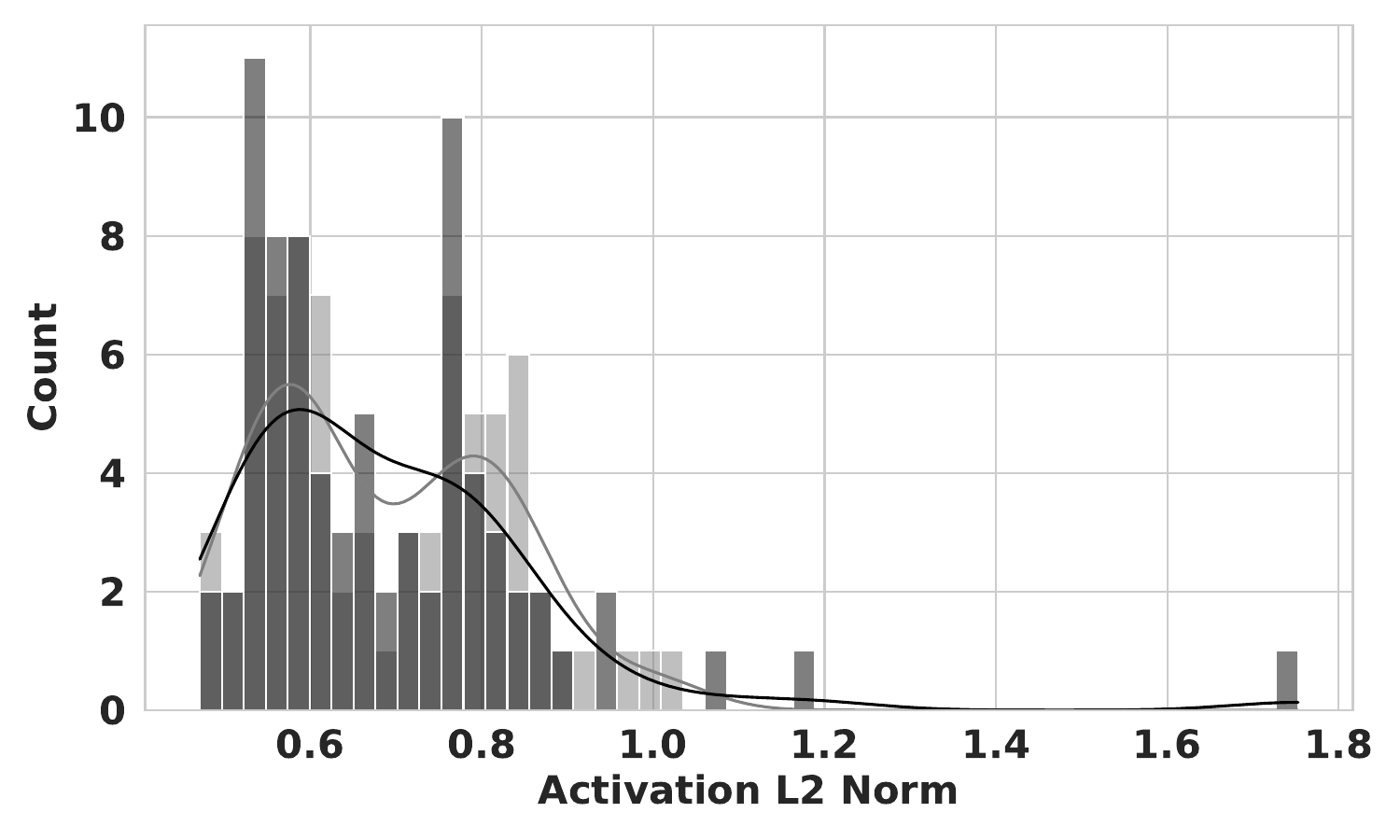}
\caption{Multi-Layer (MLP)}
\end{subfigure}
\caption{Distribution of Activation Vector Norms for Correct vs. Incorrect Predictions (for L2 Norm of Attention-Head Activations). The variation in norm ranges highlights how different probe architectures capture correctness-related signals with distinct sensitivity and scaling.}
\label{fig:activation_norm}
  \vspace{-10pt}
\end{figure*}

\begin{figure*}
\centering
\begin{subfigure}{0.325\textwidth}
\includegraphics[width=\linewidth]{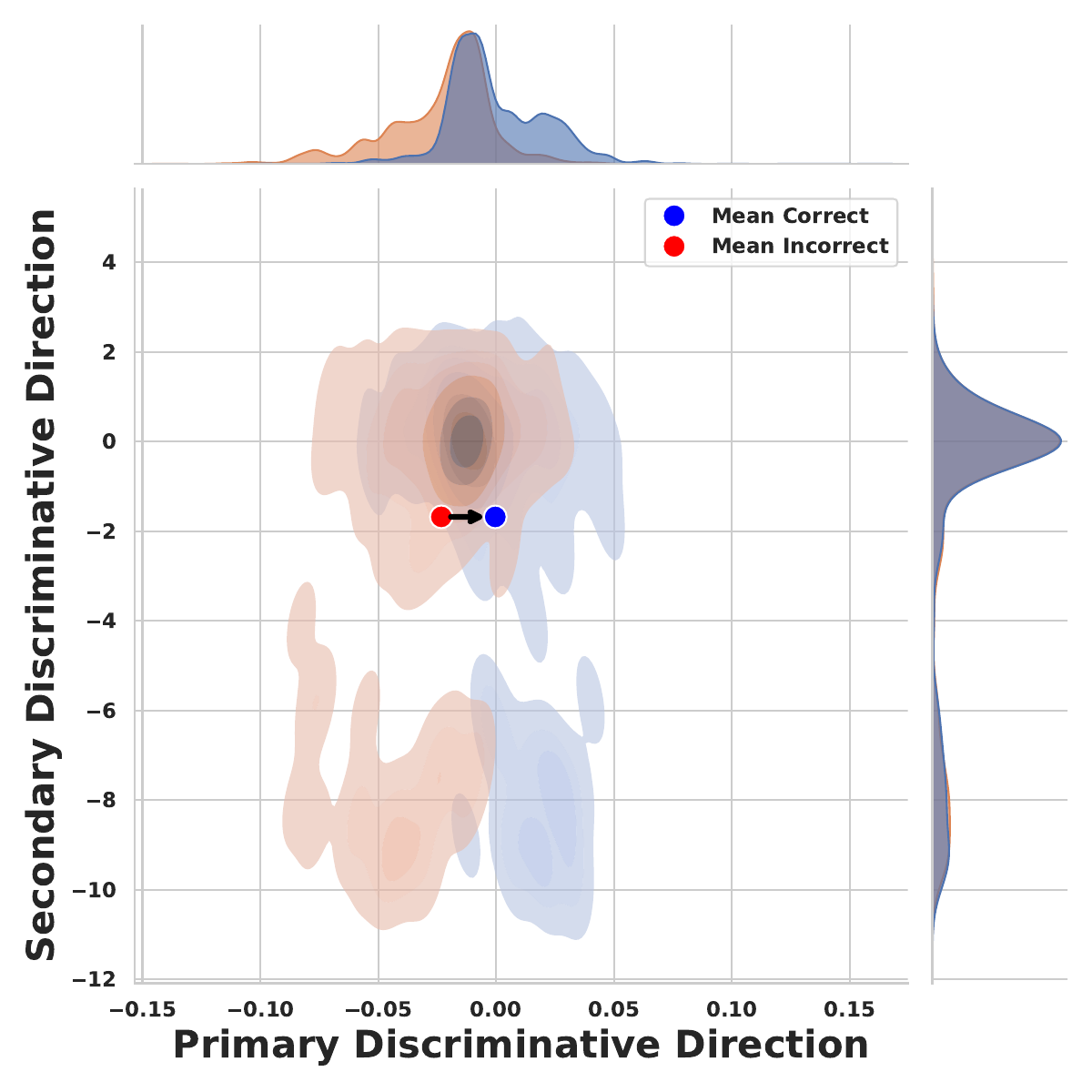}
\caption{Combinational Logic}
\end{subfigure}
\begin{subfigure}{0.325\textwidth}
\includegraphics[width=\linewidth]{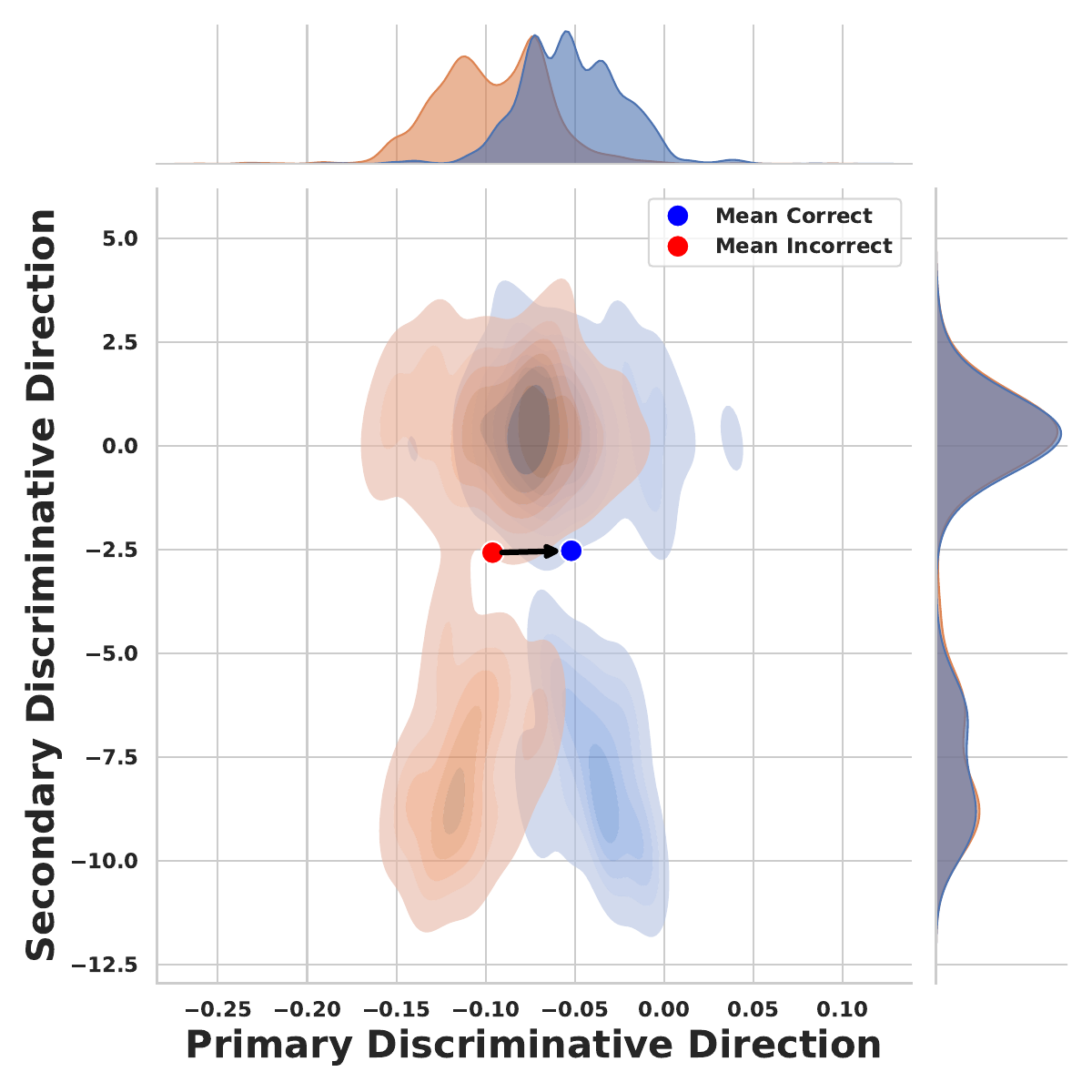}
\caption{FSM (Controller)}
\end{subfigure}
\begin{subfigure}{0.325\textwidth}
\includegraphics[width=\linewidth]{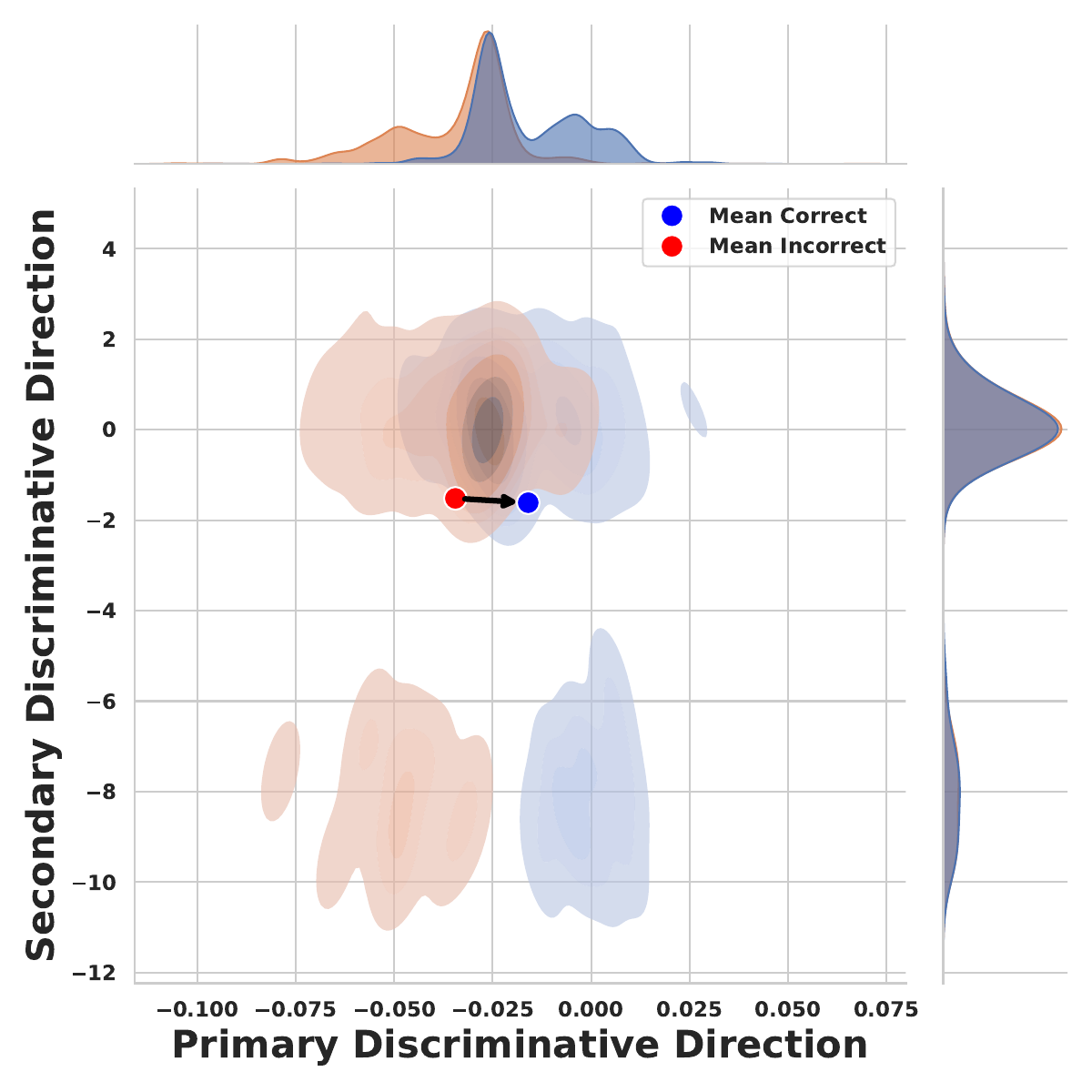}
\caption{Sequential (Datapath) Logic}
\end{subfigure}
\caption{Activation Shifts Across All Classifiers for Expert Tasks. Combinational designs show the clearest separation with distinct centroids, FSMs exhibit significant overlap, and sequential logic falls in between, reflecting the varying difficulty of correctness alignment across design categories.}
\label{fig:activation_shifts}
  \vspace{-5pt}
\end{figure*}

To analyze how probes distinguish correct from incorrect generations, we examined activation patterns across classifier types. Figure \ref{fig:activation_norm} shows L2 norm distributions: SVC with RBF probes give the clearest separation (correct 15–25 vs. incorrect 5–15), LR (linear) probes show overlap but a noticeable shift, and multi-layer (MLP) display multi-peak patterns, indicating nuanced non-linear representations. Figure \ref{fig:activation_shifts} projects discriminative directions across hardware experts. For combinational logic, correct and incorrect samples separate clearly with distinct centroid positions. Sequential logic shows tighter clusters and smaller margins, matching its lower baseline performance. FSMs exhibit the most overlap, explaining their difficulty even with specialized probes. Overall, these results confirm correctness-related signals in attention activations, beyond probe-specific artifacts.

\subsection{Intervention Optimization Analysis}
\begin{figure}[!t]
  \centering
  \includegraphics[width=\columnwidth]{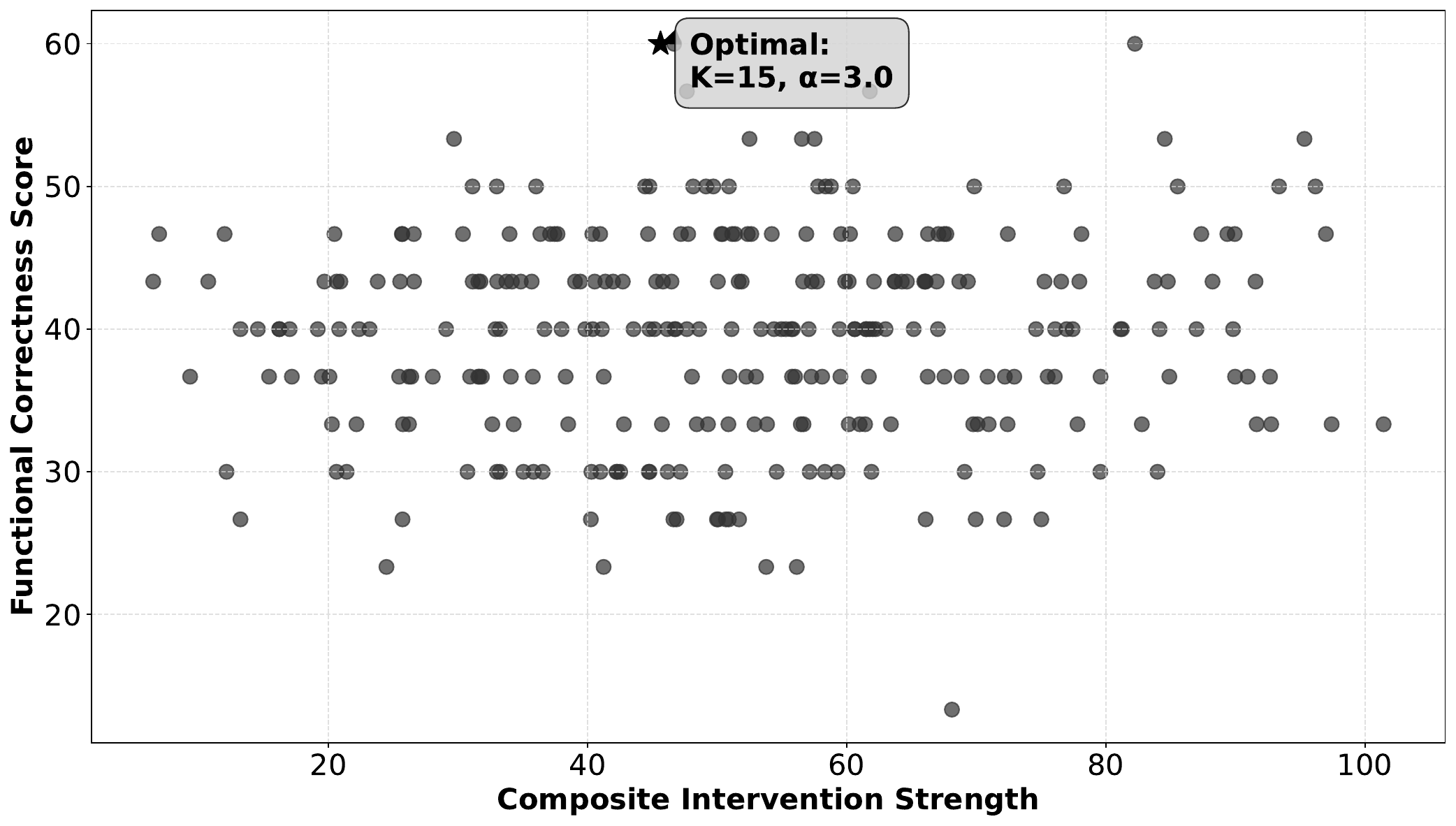}
  \caption{Intervention strength ($\alpha$) vs. correctness; peak near $K{=}15$, $\alpha{\approx}3.0$.}
  \label{fig:intervention_strength}
\end{figure}

To optimize MeltRTL’s intervention, we analyzed strength and head selection strategies. Figure \ref{fig:intervention_strength} shows a non-monotonic relationship between $\alpha$ and performance, peaking at K=15 heads. At low strength ($\alpha$ < 1.0), corrections are too weak for meaningful gains. In the optimal range ($\alpha$ = 2.5–3.5), functional correctness improves substantially. Beyond $\alpha$ = 4.0, excessive intervention destabilizes generation, as steering signals overwhelm the model’s natural patterns.

\subsection{Computational Overhead Analysis}

The computational overhead of ITI is minimal, consisting of a one-time offline setup and a lightweight per-token cost during inference. The setup phase, i.e., activation collection and probe training, is cached and does not affect generation latency. At inference, the added costs are selecting the top-$k$ heads (negligible, $O(N_p \log N_p)$) and applying interventions at $L_i$ layers, where each intervention is a vector addition of size $d_{model}$ with cost $O(d_{model})$. Formally, the cost of ITI is

\begin{equation}
C_{ITI} = C_{base} + |L_i|\cdot O(d_{model}), \quad 
\Delta C = |L_i|\cdot O(d_{model}),
\end{equation}
where the base model cost is $C_{base} \approx O(s \cdot d_{model}^2)$. Hence, the relative overhead could be calculated as follows:
\begin{equation}
\label{eq:cost_iti}
\frac{\Delta C}{C_{base}} \approx \frac{|L_i|}{s \cdot d_{model}},
\end{equation}

Equation \ref{eq:cost_iti} shows that ITI adds only simple vector additions compared to the dominant quadratic operations of transformers, leading to negligible practical latency. Table \ref{tab:overhead} provides empirical measurements of the timing overhead, showing that MeltRTL adds only 3 seconds per sample (RTL) generation.

\begin{table}[t]
\centering
\small
\setlength\tabcolsep{10pt}
\caption{Computational Overhead Analysis}
\label{tab:overhead}
\begin{tabular}{@{} p{4cm} p{1cm} p{1cm} @{}}
\toprule
\textbf{Metric} & \textbf{Base} & \textbf{MeltRTL} \\
\cmidrule(r){1-1} \cmidrule(r){2-2} \cmidrule(r){3-3}
Avg. Time / sample (s) & 11.27 & 14.27 \\
\cmidrule(r){1-1} \cmidrule(r){2-2} \cmidrule(r){3-3}
Std. Dev (s)           &  5.36 &  5.83 \\
\bottomrule
\end{tabular}
\end{table}

\section{Conclusion}

This paper introduced MeltRTL, the first framework to enhance RTL code generation through inference-time interventions at the representation level, without retraining or fine-tuning LLMs. By combining probe-guided head selection with a multi-expert steering mechanism, MeltRTL achieves substantial gains in synthesizability and functional correctness over strong base models (functional correctness from 45.3\% to 60.0\% while raising synthesizability from 85.3\% to 96.0\%), with only modest computational overhead.  By demonstrating that internal representations of LLMs can be effectively steered at inference time, MeltRTL establishes a new paradigm for LLM-based RTL generation, which (orthogonally) complements existing scaling, fine-tuning, and agentic strategies.

\bibliographystyle{IEEEtran}
\bibliography{references}

\end{document}